\documentclass{article}
\usepackage{amsfonts,amssymb,amsbsy,amsmath,amsthm,enumerate,verbatim}

\def\beq{\begin{equation}}
\def\eeq{\end{equation}}
\def\qed{\ {\vrule height5pt depth0pt width5pt}\ }

\newcommand{\R}{\mathbb{R}}

\newcommand{\C}{\mathbb{C}}

\newtheorem{theorem}{Theorem}[section]
\newtheorem{lemma}{Lemma}[section]
\newenvironment{Myproof}{\textbf{Proof.} \\ }{\qed \\ }

\begin{document}


\begin{center}
{\Large \bf Sub-linear capacity scaling for multi-path channel models}
\end{center}

\medskip
\begin{center}
{\bf F.~Bentosela$^1$, E.~Soccorsi$^1$}
\end{center}
\footnotetext[1]{Aix-Marseille Universit\'e, CNRS, CPT, UMR 7332, 13288 Marseille, France.}
\medskip

\begin{abstract}
The theoretic capacity of a communication system constituted of several transmitting/receiving elements is determined by the singular values of its transfer matrix. Results based on an independent identically distributed channel model, representing an idealized rich propagation environment, state that the capacity is directly proportional to the number of antennas. Nevertheless there is growing experimental evidence that the capacity gain can at best scale at a sub-linear rate with the system size.
In this paper, we show under appropriate assumptions on the transfer matrix of the system, that the theoretic information-capacity of multi-antenna systems is upper bounded by a sub-linear function of the number of transmitting/receiving links.
\end{abstract}



\section{Introduction}

Some wireless telecommunication systems are made of several antennas working simultaneously both at the transmit and the receive link sides. This technology named by the acronym MIMO for multiple-input-multiple-output, aims to increase the capacity of the system, i.e. the throughput of information, expressed in bits per second, being transmitted without error for a given frequency band width.

\subsection{Capacity}
An exact expression of the capacity of MIMO systems, generalizing the Shannon capacity of SISO (for single-input-single -ouput) systems has been derived by G. Foschini and M. Gans in \cite{Fos} when both transmit (Tx) and received (Rx) signals are harmonic signals, i.e. signals associated to a fixed frequency. In this case, the transmit signal at the $j^{\rm \small th}$ Tx antenna, $j=1,\ldots,N$, for $N \in {\mathbb N}^*$, together with the received signal at the $i^{\rm \small th}$ Rx antenna, $i=1,\ldots,M$, for $M \in {\mathbb N}^*$, are simply described by their respective complex amplitudes $t_j$ and $r_i$.
In presence of noise $n=(n_1,\ldots,n_M)^t \in \C^M$, the received signals $r=(r_1,\ldots,r_M)^t \in \C^M$ are related to the transmit signals $t=(t_1,\ldots,t_N)^t \in \C^N$ through the identity
\beq
\label{s0}
r = H t + n,
\eeq
where $H \in M_{M,N}({\mathbb C})$ is the transfer matrix of the system, and $M_{M,N}({\mathbb C})$ denotes the
set of $M$-by-$N$ matrices with complex elements. When $M=N$ we write $M_{N}({\mathbb C})$ instead of $M_{N,N}({\mathbb C})$.

If $n$ is a symmetric Gaussian noise with covariance matrix proportional to the identity, i.e. ${\mathbb E}(n n^*)=\nu_0 {\mathbb I}_M$ for some $\nu_0>0$, the Foschini-Gans capacity $C_M$ of the system is given by \cite{Fos} as
\beq
\label{icapa1}
C_M := \log_2 \det \left( 1 + \frac{E_S}{\nu_0 M} H H^* \right),
\eeq
where $E_S$ denotes the total power sent by the emitters. Setting $\kappa:= E_S \slash \nu_0$, $C_M$ can be expressed in terms of the singular values $\{ \mu_i \}_{i=1}^M$ of $H$ as
\beq
\label{icapa2}
C_M = \sum_{i=1}^M \log_2 \left( 1 + \frac{\kappa}{M} \mu_i^2 \right).
\end{equation}
Hence the MIMO capacity of the system is the sum of the Shannon capacities of $M$ individual SISO channels with respective power gain $\mu_i^2$, for $i=1,\ldots,M$ (see \cite{Pau, ElZ}).

\subsection{Transfer matrix modeling}
The transfer matrix of the system is fully determined by the spatial position of the antennas (in the present paper we consider one dimensional uniform linear arrays of $M=N \in {\mathbb N}^*$ antennas at both transmitter and receiver) and the scattering properties of
the propagation medium. Unfortunately there is no effective method describing the structure of $H$ in a realistic rich scattering environment. Moreover there is only a very small number of experimental or numerical data of transfer matrices available. For all these reasons several types of theoretical models for $H$ have been developed in both the physical and the mathematical literatures.

Based on the spectral theory of random matrices (see e.g. \cite{Mar, Kho, Sil, Pas, Hac1}) many capacity calculations (see \cite{Fos, Tel, Shu, Chu, Bur, Mul, Bur1, Smi}) are carried out using a priori probabilistic assumptions on the transfer matrix. Most of these {\it probabilistic models} (\cite{Fos, Tel, Bur, Mul, Bur1}) assume that $H$ consists of independent, identically distributed Gaussian random variables. These independent fading models describe a rich idealized scattering environment whose capacity gain turns out to be directly proportional to the number of transmit/receive antenna elements. Further, for independent non-identically distributed random entries, the results of \cite{Hac1} indicate that the theoretic information capacity remains asymptotically proportional to the number of antennas. Similarly \cite{Shu, Chu} show that the correlated fading capacity increases linearly with the number of antennas, but less rapidly than in independent fading.

Another approach is to define $H$ by modeling the scattering characteristics of the propagation channels. The corresponding models, based on the geometrical optics approximation and the derived ray tracing theory, are defined by a set of scattering paths $\mathcal{P}$ corresponding to scatterers distributed within the propagation medium. In these {\it scattering models}, the transmitter and receiver are coupled via propagation along
the path $p \in \mathcal{P}$ with $\Omega_{T,p}$ and $\Omega_{R,p}$ as the spatial angles seen by transmitter and receiver, and $\beta_p(\Omega_{R,p},\Omega_{T,p})$ as the corresponding fading gain. The total gain $h_{i,j}$, where $h_{i,j}$ denotes the element in the $i^{\small \rm th}$ row and $j^{\small \rm th}$ column of $H$, for the wavelength of propagation $\lambda$ is defined (see e.g. \cite{ElZ, Say, Fle1, Fle2, Ugh}) as
\beq
\label{g1}
h_{i,j} = \sum_{p \in \mathcal{P}} \beta_p(\Omega_{R,p},\Omega_{T,p}) {\rm e}^{i \frac{2 \pi}{\lambda} \langle \Omega_{T,p}, x_{T,j} \rangle} {\rm e}^{i \frac{2 \pi}{\lambda} \langle \Omega_{R,p}, x_{R,i} \rangle},
\eeq
$x_{T,j}$ and $x_{R,i}$ being the respective positions of the $j^{\rm th}$ transmitter and the $i^{\rm th}$ receiver.
The transfer matrix of the system (\ref{g1}) is thus directly defined from the transmitter and receiver positions together with the physical characteristics of the propagation medium.

\subsection{Framework and main result}
In this paper we adopt the scattering model point of view by imposing structure on $H$, based on simulation results obtained with an efficient 3D ray propagation model in some reference urban outdoor environment, for one dimensional uniform linear arrays of $M$ antennas at both transmit and receive links. More precisely, computations of the transfer matrix are carried out for numerous values of $M \in {\mathbb N}^*$, using the (\ref{g1})-based propagation model GRIMM developed by France Telecom (see \cite{Ros1, Ros2, Ros3}).
A careful analysis of these data shows that the fading matrix $F$ of the system,
\beq
\label{g2}
F := M^{-2} U_M H H^* U_M^{-1} \in M_M({\mathbb C}),
\eeq
where $U_M$ denotes the unitary change of basis matrix from the canonical basis of $\C^M$ to the Fourier basis $\{ \Phi_k \}_{k=1}^M$, defined in \cite{Say} as the virtual channel representation,
$$ \Phi_k := \frac{1}{M^{1 \slash 2}}(1,e^{i 2 \pi k \slash M},\ldots,e^{i 2 \pi (M-1)k \slash M})^t,\ k=1,\ldots,M, $$
has the two following properties:
\begin{enumerate}
\item[(i)] for all $i=1,\ldots,M-1$, the off-diagonal terms $f_{k,j}$, $j>k \geq i$, of $F$, are small as compared with the diagonal term $f_{i,i}$;
\item[(ii)] $\{ f_{i,i} \}_{i=1}^M$ can be reordered into a decreasing sequence.
\end{enumerate}
To avoid the inadequate expense of the size of this article we refer to \cite{Ben} for both the justification and the interpretation of (i)-(ii). These two properties (or, more exactly, their corresponding appropriate mathematical statement, formulated as assumptions (A1)-(A2) in Section 2) provide useful spectral information on $F$. Namely they enable a precise localization of the eigenvalues of $F$, involving that the capacity $C_M$ of the system is upper bounded by a sub-linear function of the system size, $M$. This is the main result of this paper.

\subsection{Contents}
The paper is organized as follows. In Section 2 we introduce some basic notations and auxiliary results, state the assumptions (A1)-(A2) made on the fading matrix $F$, formulate the main results of this paper and briefly comment on them. Section 3 contains the proof, based on an appropriate block-decomposition of $F$, of these results.
The main technical estimate needed to conclude the proof in Section 3 (this estimate actually holds true for any Hermitian matrix independently of the assumptions (A1)-(A2) intrinsic to the model of Section 2) is given in Section 3.2.

\section{Capacity scaling}
\subsection{Notations and settings}
In this section we introduce some notations used throughout the article and recall basic auxiliary estimates
needed in the proofs.

Let $A=(a_{i,j})_{1 \leq i,j \leq n} \in M_n(\C)$, $n \geq 1$, be an Hermitian matrix. We denote by $P_I(A)$ the spectral projection of $A$ corresponding to the open interval $I \subset \R$ and set
\[ N(x;A) := {\rm rank}\ P_{(x,+\infty)}(A),\ x \in \R. \]
Otherwise stated $N(x;A)$ denotes the number of eigenvalues of $A$ counted with the multiplicities and greater than $x$. $N(.;A)$ is called the eigenvalue counting function of $A$.

In the sequel we write $\sigma(A)$ (resp. $\rho(A):=\C - \sigma(A)$) the spectrum (resp. resolvent set) of $A$, and $A_D$ (resp. $A_O:=A-A_D$) the diagonal (resp. off-diagonal) part of $A$.

Further, noting $\| A \|$ the matrix norm of $A$ associated to the Hermitian norm in $\C^n$,
we have (see (I.4.14) and (I.4.16) in \cite{Kat}) that
\beq
\label{norest1}
\| A \| \leq \max_{i=1,\ldots,n} \tau_i(A),
\eeq
where
\beq
\label{norest2}
\tau_i(A):= \sum_{j=1,\ldots,n} |a_{i,j}|.
\eeq

Finally we recall from the standard
perturbation theory (see \cite{Kat}) that the eigenvalues $\{ \mu_i(A) \}_{i=1}^n$ of $A$ may be labeled in such a way that the Bauer-Fike's Theorem holds true:
\beq
\label{BF}
| \mu_i(A) - a_{ii} | \leq \| A_O \|,\ i=1,\ldots,n.
\eeq

\subsection{Structure of the fading matrix}

Evidently, the diagonal terms $f_i := f_{i,i}$, for $i=1,\ldots,M$, of the matrix $F$ defined in (\ref{g2}) are nonnegative, and we may assume without loss of generality that they are arranged in decreasing order:
\beq
\label{decorder}
0 \leq f_{M} \leq \ldots \leq f_{2} \leq f_{1}.
\eeq
In light of the properties (i)-(ii) mentioned in Section 1.3, we make the two following assumptions on the fading matrix $F$.

\paragraph{Assumption 1.}
Our first assumption expresses for each $i=1,\ldots,M-1$, that the total weight of the off-diagonal terms $f_{k,j}$, for $j>k \geq i$, is bounded, up to a multiplicative constant $\alpha>0$, by the diagonal element $f_i$:
$$ \exists \alpha>0,\ \forall M \geq 1,\ \sum_{j >k \geq i} | f_{k,j} | \leq \alpha f_i,\ i=1,\ldots,M-1.\ \ \ \ \ \ \mathrm{\bf (A1)}$$
As proved in Lemma \ref{lm-normat} below, (A1) yields that the $(M-i_0)$-square matrix
\beq
\label{defio}
\tilde{F}^{(i_0)}:=\left( f_{i,j} \right)_{i_0+1 \leq i,j \leq M} \in M_{M-i_0}({\mathbb C}),\ i_0=0,\ldots,M-1,
\eeq
obtained from $F$ by suppressing its $i_0$ first rows and columns, satisfies:
\beq
\label{A1bis}
\| \tilde{F}^{(i_0)}_O \| \leq \alpha f_{i_0+1},\ i_0=0,\ldots,M-1.
\eeq

\paragraph{Assumption 2.}
Further, we impose that the function $i \mapsto f_i$ decreases sufficiently fast with $i$ on $\{1,\ldots,M \}$.
More precisely we consider a power-like decay and require that the power decay rate be greater than one:
$$ \exists (f_+,\gamma) \in (0,+\infty) \times (1,+\infty),\ \forall M \geq 1,\ f_i \leq f_+ i^{-\gamma},\ i=1,\ldots,M.\ \ \ \ \ \ \mathrm{\bf (A2)}$$

Notice for further reference that (A2) entails
\beq
\label{ca2}
\sharp \{ i=1,\ldots,M\ {\rm s.t.}\ f_i \in (x,+\infty) \} \leq \min \left( M, f_+^{\gamma^{-1}} x^{-\gamma^{-1}} \right),\ x \in (0,\infty),
\eeq
where $\sharp S$ denotes the cardinality of any subset $S$ of ${\mathbb N}$.

\subsection{Statement of the main results}
By (\ref{g2}), the fading matrix $F$ is a nonnegative Hermitian matrix. Thus the eigenvalue counting function $N(x;F)$ of $F$ is a non increasing right-continuous function of $x \in \R$,
such that
\begin{equation}
\label{densta1}
N(x;F) =M,\ x < 0,\ \mbox{and}\ N(x;F) = 0,\ x \geq \rho,
\end{equation}
where $\rho:=\max_{i=1,\ldots,M} \lambda_i$ is the spectral radius of $F$.
The first result of this paper is a convenient upper bound on $N(x;F)$:

\begin{theorem}
\label{thm-IDSpol}
Let $M \in {\mathbb N}^*$ and $F$ satisfy (A1)-(A2).
Then we have
\beq
\label{ecf}
N(x;F) \leq \min \left( M, \rho_+^{\gamma^{-1}} x^{-\gamma^{-1}} \right),\ x \in (0,+\infty),
\eeq
where
\beq
\label{sr}
\rho_+:=(1+\alpha) f_+ \geq \rho.
\eeq
\end{theorem}

As a corollary we obtain under the same conditions that for $\gamma>1$, the Foschini-Gans capacity growths sub-linearly with the system size $M$:

\begin{theorem}
\label{thm-capa}
Let $F$ be as in Theorem \ref{thm-IDSpol}.
Then the capacity $C_M$ of the system is bounded as
$$ C_M \leq \frac{(\kappa \rho_+)^{\gamma^{-1}}}{\ln 2} M^{\gamma^{-1}} \left( \frac{\gamma}{\gamma-1}+\ln(1+\kappa \rho_+ M) \right), $$
the constant $\rho_+$ being defined by (\ref{sr}).
\end{theorem}

Therefore the capacity $C_M$ growths at most like $M^{\gamma^{-1}} \ln M$. For $\gamma>1$, it is thus upper bounded by a sub-linear function of $M$, at least for $M$ sufficiently large.
This behavior is different from the one predicted by \cite{Fos, Tel, Shu, Chu, Bur, Mul, Bur1} for probabilistic models, where a point-to-point link utilizing $M$ transmitting and receiving antennas can achieve a capacity as high as $M$ times that of a single-antenna link. Nevertheless it is in accordance with the results of \cite{Bur1, Rag}, where
similar sub-linear capacity scalings are derived for multi-path scattering models. Moreover the upper bound given in Theorem \ref{thm-capa} is corroborated by the simulation results obtained in \cite{Ger} and the experimental evidence of \cite{Mol, Wal}. This indicates that the assumptions (A1)-(A2) can be considered a valuable alternative to pure probabilistic models, in the study of MIMO-systems theoretic capacity.

\subsection{Comments}
In view of Theorems \ref{thm-IDSpol}-\ref{thm-capa}, we make the three following remarks:
\begin{enumerate}
\item[(i)] In this paper, in common with works such as \cite{Fos, Tel}, but unlike \cite{Mul, Bur1}, we do not assume a normalization which ensures that the total receive power is the same as the total transmit power, that is
$$ \sum_{i,j=1}^M | h_{i,j} |^2 = M. $$
Such a normalization condition would imply $\| H \| \leq M^{1 \slash 2}$ and hence $\| F \| \leq M^{-1}$ by (\ref{g2}), enabling us to carry out all the computations of Section 3 with the constant $\rho_+$, defined in
(\ref{normat}), equal to $M^{-1}$. Nevertheless it is quite easy to check that this would not significantly change the conclusions of Theorems \ref{thm-IDSpol}-\ref{thm-capa}.
\item[(ii)] As already mentioned in Section 2.2 (see (\ref{A1bis})), assumption (A1) implies:
$$ \exists \alpha>0,\ \forall M \geq 1,\ \| \tilde{F}^{(i_0)}_O \| \leq \alpha f_i,\ i=1,\ldots,M-1,\ \ \ \ \ \ \mathrm{\bf (A1')}$$
the matrix $\tilde{F}^{(i_0)}$, $i_0=1,\ldots,M-1$, being defined in (\ref{defio}).
Actually it is not hard to check from the proofs of Section 3 that Theorems \ref{thm-IDSpol}-\ref{thm-capa} remain true by substituting the weaker (but less explicit) condition (A1') for (A1).
\item[(iii)] Similarly, assumption (A2) is a particular case of the more general condition
$$ \left\{ \begin{array}{c} \mathrm{There\ exists}\ f : [1,+\infty) \rightarrow {\mathbb R}_+,\ \mathrm{continuous\ and\ decreasing,} \\
\mathrm{s.t.}\ \forall M \geq 1,\ f_i \leq f(i),\ i=1,\ldots,M,  \end{array} \right.\ \mathrm{\bf (A2')}$$
and it is easy to see that Theorem \ref{thm-IDSpol} generalizes to
$$N(x;F) \leq \min \left( M, f^{-1}((1+\alpha)^{-1} x) \right),\ x \in (0,+\infty), $$
where $f^{-1}$ denotes the inverse function of $f$, for every $F$ satisfying (A1)-(A2') (or (A1')-(A2'), according to point (ii)).

For instance, if we strengthen (A2) by taking $f(x)=f_+ e^{-\gamma (x-1)}$, for some $(f_+,\gamma) \in (0,+\infty)^2$, and thus imposing that the diagonal elements of $F$ decrease exponentially fast,
$$ \forall M \geq 1,\ f_i \leq f_+ e^{-\gamma (i-1)},\ i=1,\ldots,M, $$
then the capacity of the corresponding system is upper-bounded by a polynomial function in $\ln M$:
$$ C_M \leq \frac{\gamma^{-1} + \ln(1+\kappa \rho_+ M) + \gamma^{-1}\ln(1+\kappa \rho_+ M)^2}{\ln 2}. $$
This estimate can be easily obtained by mimicking the proof of Theorem \ref{thm-capa}.
\end{enumerate}

\section{Analysis of the capacity}

\subsection{Proof of Theorem \ref{thm-IDSpol}}
The proof of Theorem \ref{thm-IDSpol} consists of the two following lemmas.

\begin{lemma}
\label{lm-normat}
Assume (A1). Then for all $M \in {\mathbb N}^*$ and $i_0=0,\ldots,M-1$, the $(M-i_0)$-square matrix
$\tilde{F}^{(i_0)}$ defined in (\ref{defio}) satisfies
\beq
\label{normat}
\| \tilde{F}^{(i_0)}_O \| \leq \alpha f_{i_0+1}\ {\rm and}\ \| \tilde{F}^{(i_0)} \| \leq (\alpha+1) f_{i_0+1}.
\eeq
\end{lemma}
\begin{Myproof}
For all $i=i_0+1,\ldots,M$, we have
\beq
\label{z1}
\tau_i(\tilde{F}^{(i_0)}_O) = \sum_{j=i_0+1,\ldots,i-1} |f_{i,j}| + \sum_{j=i+1,\ldots,M} | f_{i,j} |,
\eeq
according to (\ref{norest2}), the first (resp. second) term in the righthand side of (\ref{z1}) being taken equal to zero if $i=i_0+1$ (resp. $i=M$). The matrix $F$ being Hermitian symmetric by (\ref{g2}), (\ref{z1}) becomes
$$
\tau_i(\tilde{F}^{(i_0)}_O) = \sum_{j=i_0+1,\ldots,i-1} |f_{j,i}| + \sum_{j=i+1,\ldots,M} | f_{i,j} |
\leq \sum_{j>k \geq i_0+1} | f_{k,j}|,
$$
whence
\beq
\label{z2}
\tau_i(\tilde{F}^{(i_0)}_O) \leq \alpha f_{i_0+1},\ i=i_0+1,\ldots,M.
\eeq
from (A1). This combined with (\ref{norest2}) and (\ref{decorder}) yields
\beq
\label{z3}
\tau_i(\tilde{F}^{(i_0)}) = \tau_i(\tilde{F}^{(i_0)}_O) + f_{i} \leq (\alpha+1) f_{i_0+1},\ i=i_0+1,\ldots,M.
\eeq
Now the first (resp. second) part of (\ref{normat}) follows from (\ref{norest1}) and (\ref{z2}) (resp. (\ref{norest1}) and (\ref{z3})).
\end{Myproof}

\begin{lemma}
\label{lm-densta}
Let $M$ be in ${\mathbb N}^*$. If the fading matrix $F$ satisfies $\mathrm{(A1)}$ then it holds true that
$$ N(x;F) \leq N((1+\alpha)^{-1} x;F_D),$$
for all $x \in [0,+\infty)$.
\end{lemma}
\begin{Myproof}
For all $x \in [0,+\infty)$ set $n(x)=N((1+\alpha)^{-1} x;F_D)$, so that $n(x) \in \{0,1,\ldots,M \}$. The result being obviously true if $n(x)=M$, we assume that $n(x)<M$, which entails
\beq
\label{cl}
f_{n(x)+1} \leq \frac{x}{1+\alpha}.
\eeq
Further $F$ decomposes uniquely into
$$F := \left( \begin{tabular}{c|c} $A(x)$ & $C(x)$ \\ \hline $C(x)^*$ & $B(x)$ \end{tabular} \right),
$$
where $B(x)$ is the matrix $\tilde{F}^{(n(x))} \in M_{M-n(x)}({\mathbb C})$ defined by (\ref{defio}).
Since
$$ \| B_O(x) \| \leq \alpha f_{n(x)+1}, $$
by the first part of (\ref{normat}) in Lemma \ref{lm-normat}, then each eigenvalue $\beta_j$, $j=n(x)+1,\ldots,M$, of $B(x)$, satisfies
$$ \beta_j \leq f_j + \| B_O(x) \| \leq (1+\alpha) f_{n(x)+1} \leq x, $$
according to (\ref{BF}), (\ref{decorder}) and (\ref{cl}). This yields $N(x;B(x))=0$, so the result follows from the inequality
\beq
\label{mte}
N(x;F) \leq n(x) + N(x;B(x)),
\eeq
whose proof is postponed to Section 3.2 below.
\end{Myproof}

Bearing in mind that
$$ N(x;F_D) = \sharp \{ i=1,\ldots,M\ {\rm s.t.}\ f_i \in (x,+\infty) \},\ x \in (0,+\infty),$$
Theorem \ref{thm-IDSpol} follows immediately from this, (\ref{ca2}) and Lemma \ref{lm-densta}.

\subsection{Proof of inequality {\bf \eqref{mte}}}
Inequality \eqref{mte} follows from Lemma \ref{l1} below, which is in turn a consequence of a result known as the Cauchy interlacing Theorem (see \cite{Car}[Theorem 3(i)]).
\begin{lemma}
\label{l1}
Let $n_A$, $n_B$ be positive integers. Given $A = A^* \in M_{n_A}(\C)$, $B = B^* \in M_{n_B}(\C)$ and $C \in M_{n_A,n_B}(\C)$, let $M$ be the partitioned $(n_A+n_B) \times (n_A+n_B)$ matrix defined by
$$
M := \left( \begin{tabular}{c|c} $A$ & $C$ \\ \hline $C^*$ & $B$ \end{tabular} \right).
$$
Then we have
\beq
\label{zz1}
 N(x;M) \leq n_A + N(x;B)
\eeq
for all real $x$.
\end{lemma}
\begin{Myproof}
Given an Hermitian matrix $H$, and a real number $x$, let $\tilde{N}(x;H)$ denote the number of eigenvalues, counted with multiplicity, less than or equal to $x$. With this notation we obviously have
$$ N(x;M) + \tilde{N}(x;M)=n_A+n_B\ {\rm and}\ N(x;B) + \tilde{N}(x;B) = n_B, $$
for all real $x$, whence \eqref{zz1} is equivalent to the inequality
\beq
\label{zz2}
\tilde{N}(x;B) \leq \tilde{N}(x;M),\ x \in \R.
\eeq
Thus it suffices to establish \eqref{zz2}.
To this purpose the eigenvalues of $M$ are written as
$$ \lambda_1 \leq \lambda_2 \leq \ldots \leq \lambda_{n_A+n_B} $$
and repeated according to multiplicity. Similarly we write for the eigenvalues of $B$
$$ \beta_1 \leq \beta_2 \leq \ldots \leq \beta_{n_B}. $$
The Cauchy interlacing Theorem states that we have
$$ \lambda_j \leq \beta_j \leq \lambda_{n_A+j},\ j=1,2,\ldots,n_B.  $$
Evidently the lefthand inequality yields  \eqref{zz2}, proving the result.
\end{Myproof}

\subsection{Proof of Theorem \ref{thm-capa}}
In light of (\ref{icapa1})-(\ref{icapa2}), the Foschini-Gans capacity of the system associated to the fading matrix $F_M$ can be expressed in terms of the eigenvalues $\{ \lambda_i \}_{i=1}^{M}$ of $F$, as
\begin{equation}
\label{capa1}
C_M = \sum_{i=1}^M \log_2 \left( 1 +  \kappa M \lambda_i \right).
\end{equation}
The remaining part of this section involves relating the capacity $C_M$ to
the eigenvalue counting function $N(.;F)$ of the matrix $F$.
To this end, we use the distributional equality
$dN(x;F) = -\sum_{i=1}^M \delta(x-\lambda_i)$ to rewrite (\ref{capa1}) as
\begin{equation}
\label{capa2}
C_M 
= \int_{0}^{\rho_+} \log_2 \left( 1 + \kappa M  x \right) (-dN(x;F)).
\end{equation}
Integrating by parts in (\ref{capa2}) and bearing in mind (\ref{densta1}), we obtain
$$
C_M = \frac{\kappa M}{\ln 2} \int_{0}^{\rho_+} \frac{N(x;F)}{1 + \kappa M x} dx,
$$
whence
$$
C_M= \frac{1}{\ln 2} \int_{0}^{\kappa \rho_+ M} \frac{N((\kappa M)^{-1}x;F)}{1 +x} dx
$$
by an obvious change of integration variable.
This combined with Theorem \ref{thm-IDSpol} yields
\beq
\label{capa3}
C_M \leq \frac{(\kappa \rho_+)^{\gamma^{-1}}}{\ln 2} M^{\gamma^{-1}} \int_{0}^{\kappa \rho_+ M} \frac{x^{-\gamma^{-1}}}{1 + x} dx.
\eeq
If $\kappa \rho_+ M > 1$ then the integral domain in the righthand side of (\ref{capa3}) can be partitioned into $(0,1)$ and $(1, \kappa \rho_+ M)$. Since $\gamma>1$, the integral over $(0,1)$ is easily bounded by
$\gamma \slash (\gamma - 1)$, while the one over $(1, \kappa \rho_+ M)$ is majorized by $\ln (1+\kappa \rho_+ M)$.
In the case where $\kappa \rho_+ M \leq 1$ then the integral in the righthand side of (\ref{capa3}) is evidently bounded by
$\gamma \slash (\gamma - 1)$.
This completes the proof.

\medskip
\noindent{\bf Acknowledgments.}
The authors
are greatly indebted to the anonymous referee of this paper for the considerable simplification of their original proof of Lemma \ref{l1} in Section 3.2.

\end{document}